\RequirePackage{fix-cm}
\documentclass[twocolumn,epjc3,showpacs,preprintnumbers,amsmath,amssymb]{svjour3}
\smartqed  % flush right qed marks, e.g. at end of proof
\RequirePackage{graphicx}
\RequirePackage[numbers,sort&compress]{natbib}
\usepackage{graphicx,amsmath,amsfonts,latexsym,amssymb,graphics,epsfig,subfigure,color,makeidx,hyperref}
\usepackage{multirow}
\usepackage{mathtools, cuted}

\journalname{Eur. Phys. J. C}

\newcommand\beq{\begin{equation}}
\newcommand\eeq{\end{equation}}
\newcommand\beqn{\begin{eqnarray}}
\newcommand\eeqn{\end{eqnarray}}
\newcommand\bal{\begin{align}}
\newcommand\eal{\end{align}}

\begin{document}

\allowdisplaybreaks

\title{Shortcut in codimension-2 brane cosmology in light of GW170817}
%{Constraining the AdS$_{6}$ radius with GW170817/GRB 170817A}

\author{Zi-Chao Lin\thanksref{addr1,addr2,e1} \and Hao Yu\thanksref{addr3,e2} \and Yu-Xiao Liu\thanksref{addr1,addr2,addr4,e3}
}

\thankstext{e1}{linzch18@lzu.edu.cn}
\thankstext{e2}{yuhaocd@cqu.edu.cn}
\thankstext{e3}{liuyx@lzu.edu.cn, corresponding author}

\institute{
Lanzhou Center for Theoretical Physics, Key Laboratory of Theoretical Physics of Gansu Province, School of Physical Science and Technology, Lanzhou University, Lanzhou 730000, China \label{addr1}
\and
Institute of Theoretical Physics $\&$ Research Center of Gravitation, Lanzhou University, Lanzhou 730000, China\label{addr2}
\and
Physics Department, Chongqing University, Chongqing 401331, China\label{addr3}
\and
Key Laboratory for Magnetism and Magnetic of the Ministry of Education, Lanzhou University, Lanzhou 730000, China\label{addr4}
 }

\maketitle

\abstract{
In this paper, our universe is regarded as a codimension-2 brane embedded in a noncompact six-dimensional Anti-de Sitter (AdS) spacetime. We derive the gravitational horizon radius on the brane under the low-energy approximation, which reflects how the extra dimensions cause the shortcut effect of gravitational waves (GWs). We also study the time delay between a GW signal and an electromagnetic (EM) wave signal in the low-redshift limit by combining with the joint observations of GW170817/GRB 170817A, which gives an upper limit to the $\text{AdS}_{6}^{~}$ radius as $\ell^{2}_{~} \lesssim 3.84\,\text{Mpc}^{2}_{~}$. For a high-redshift source, the time delay is converted into the discrepancy between the source redshift derived from the GW signal and the one derived from the EM counterpart. It is found that if one expects to detect the EM counterpart of a high-redshift GW event within a reasonable observation time, it requires a stronger constraint on the $\text{AdS}_{6}^{~}$ radius. Our research shows that the $\text{AdS}_{6}^{~}$ radius should satisfy $\ell^{2}_{~}\lesssim0.02\,\text{Mpc}^{2}_{~}$ for the DECIGO and BBO.
}

%\keywords{Black holes, critical phenomena, phase diagram}

%\pacs{04.50.Kd, 04.50.+h}

\section{Introduction}

The idea of using extra dimensions to unify Maxwell's electromagnetism and Einstein's gravity was first proposed by Kaluza and Klein (KK) one century ago~\cite{Kaluza1,Klein1,Klein2}. After several decades of development, people realized that both of the weak and strong interactions could be unified with the gravitational interaction through extra dimensions~\cite{Scherk1,Dai1,Horava1,Horowitz1} as well. But there still exists a huge hierarchy between the fundamental scales of the gravitational interaction and electro-weak interaction in the theory. To solve the hierarchy problem, Arkani-Hamed, Dimopoulos, and Dvali (ADD) constructed a well-known braneworld model, the so-called ADD model~\cite{ADD1,ADD2}, with large extra dimensions. Later, Randall and Sundrum (RS) developed the ADD model and successfully solved the hierarchy problem with a warped geometry in their RS-1 model~\cite{RS1}. For almost a century, exploring extra dimensions has always been an important topic on understanding the nature of our universe. A manifest feature of higher-dimensional theories is that they predict the existence of massive KK particles beyond the Standard Model of particle physics. In the braneworld theory, these particles can propagate in the bulk and participate in the interactions on the brane at high energy. So in the past several decades, people  seeking the evidence of extra dimensions mainly attempted to create these KK particles through particle collisions. Recently, with the progress of gravitational wave (GW) observations, some researchers have devoted into the study of GWs in the braneworld theory.

If KK gravitons could be created during the merger of a binary system, they will leak into the bulk~\cite{Pardo1,Abbott1,Dvali1,Deffayet1,Corman1,Hernandez:2021qfp}. It leads to an extra dispersion of the GW energy comparing with general relativity (GR), and makes the amplitude of the GW weaker than the one predicted in GR. As a result, the GW and electromagnetic wave (EMW) observations will give different source luminosity distances for a same GW event, such as GW170817~\cite{Abbott2,Abbott3,Abbott4}. And one can use the difference to limit the number of extra dimensions. In addition, if KK gravitons could be created in the interior of a star, there could be a novel energy-loss channel during the collapse of a star~\cite{Sakstein1}. Consequently, it allows the existence of the black hole whose mass is within the black hole mass gap predicted in the standard stellar evolution theory. Therefore, the mass of the binary components in the event GW190521 is well explained by this novel energy-loss channel~\cite{LIGOScientific2,LIGOScientific3}. For more studies on GWs in extra-dimensional theories, one can refer to Refs.~\cite{Andriot1,Du1,Garcia-Aspeitia1,Cardoso1,Chakravarti1,Chakravarti2} and references therein. These researches show that detection of GWs is a new method to investigate extra dimensions. Inspired by these researches, we will use the shortcut effect of GWs to discuss the structure of extra dimensions of a six-dimensional Anti-de Sitter (AdS) spacetime.

The conception of shortcut was first introduced by Chung and Freese~\cite{Chung1}. They found that if a signal can leave the brane and propagate in the bulk, the time it takes to travel from one point to another one on the brane could be shorter than the time that a signal (with the same speed) confined on the brane takes~\cite{Chung1}. It means that the trajectory of the previous signal could be a shorter path that causally connects the two points. Such phenomenon is named as the shortcut effect in high-dimensional spacetime, and the corresponding trajectory of the previous signal is called a shortcut. Obviously, the difference between the two trajectories is closely related to the structure of the brane~\cite{Ishihara1}. Only if the brane has a vanishing extrinsic curvature, the two trajectories correspond to the same path. So with the time difference between the two signals on the brane, one can constrain various higher-dimensional models~\cite{Yu1,Visinelli1,Yu2,Wang2002,Lin1,Caldwell1}.

In 2017, the joint GW and EMW observations reported a GW event (GW170817) and its EM counterpart (GRB 170817A) originated from the merger of a binary neutron star in NGC 4993~\cite{Abbott2,Coulter1,Pan1,Abbott3,Abbott4}. If the binary emitted the two signals at the same time, GR requires the two signals to arrive at the earth simultaneously. However, it was found that there is a $1.74^{+0.05}_{-0.05}\,\text{s}$ time delay between their arrivals. To explain the time delay, the shortcut of GWs in five-dimensional braneworld models was discussed in Refs.~\cite{Visinelli1,Yu1,Yu2,Lin1}. In the models, GWs are allowed to propagate in the background spacetime, while EMWs are confined on a 4-brane with maximally symmetric. The result shows that the shortcut of GWs can well explain the time delay in the event GW170817/GRB 170817A. With the help of the time delay, the five-dimensional AdS radius and five-dimensional de Sitter (dS) radius are restricted to $\ell\lesssim0.535\,\text{Mpc}$ and $\ell\gtrsim2.4\times10^{3}_{~}\,\text{Tpc}$, respectively~\cite{Visinelli1,Lin1}.

For a six-dimensional model, if one extra dimension is compact, its scale could be tiny and so the correction of this compact extra dimension to the GW trajectory is usually negligible. Since the gravitational horizon radius in such a six-dimensional model approximately equals the one obtained in a five-dimensional model~\cite{Cuadros-Melgar1,Abdalla1,Abdalla2,Abdalla3}, there is no novel contribution to the time delay. In this paper, we will consider a six-dimensional model with two infinite extra spatial dimensions. Under the low-energy approximation, we calculate the gravitational horizon radius and analyze the contribution of the brane's motion to the gravitational horizon radius. Especially, those corrections could be considerable when the source redshift is large. Then, we can obtain a constraint on the $\text{AdS}_{6}^{~}$ radius with GW170817/GRB 170817A. For a high-redshift source, we can generalize the formulas and give a stronger constraint on the $\text{AdS}_{6}^{~}$ radius according to future observations.

The paper is arranged as follows. In Sec.~\ref{sec2}, we construct a six-dimensional AdS spacetime and embed a 4-brane inside the bulk. Under the low-energy approximation, the expression of the gravitational horizon radius is deduced in Sec.~\ref{sec3}. Then, Sec.~\ref{sec4} is devoted to calculating the time delay in low-redshift limit and its generalization for a high-redshift source. Our constraints on the AdS$_{6}$ radius with GW170817/GRB 170817A and predictions for future observations are presented in Sec.~\ref{sec5}. Finally, our conclusion is given in Sec.~\ref{sec6}.

\section{Background Spacetime and Brane}\label{sec2}

We start with a six-dimensional static spacetime with the metric written as follows~\cite{Kanno1,Bostock1}:
\begin{equation}\label{6metric1}
	\text{d}s^{2}_{6}=-f^{2}_{~}(R)\text{d}T^{2}_{~}+R^{2}_{~}\text{d}\psi^{2}_{~}+h^{2}_{~}(R)\text{d}R^{2}_{~}+\omega(R)\text{d}\Sigma^{2}_{3},
\end{equation}
where $T$ is the bulk time. The arguments $R$ and $\psi$ are the polar coordinates spanning on the extra 2-space. They are related to the fourth and fifth spatial dimensions through $x^{5}_{~}=R\,\text{cos}\psi$ and $x^{6}_{~}=R\,\text{sin}\psi$. We use $\text{d}\Sigma_{3}^{2}$ to represent the line element of a maximally symmetric 3-space:
\begin{equation}
	\text{d}\Sigma_{3}^{2}=\frac{1}{1-kr^{2}_{~}}\text{d}r^{2}_{~}+r^{2}_{~}(\text{d}\theta^{2}_{~}+\text{sin}^{2}_{~}\theta \text{d}\phi^{2}_{~}),
\end{equation}
where $k$ is the curvature of the 3-space. In this paper, we ignore the back-reaction of the 4-brane (our four-dimensional universe) on the background spacetime for the sake of simplicity. There is only a bulk cosmological constant $\Lambda_{6}^{~}$ in the whole spacetime. We consider the six-dimensional Einstein-Hilbert action, which depends on the
Ricci scalar of the metric (\ref{6metric1}). Combining with the bulk cosmological constant, the effective
action is given by~\cite{Cho:1991cja}
\begin{equation}
	S=\int \sqrt{-g}\text{d}^{6}_{~}x\frac{1}{16\pi G_{6}^{~}}(R^{*}_{~}-2\Lambda_{6}^{~}),
\end{equation}
where $G_{6}^{~}$ is the six-dimensional Newtonian gravitational constant. Note that we use $R^{*}_{~}$ to represent the Ricci scalar in order to distinguish it from the polar radius $R$. One can obtain the solution of the metric~\eqref{6metric1} by solving the following field equations:
\begin{subequations}\label{fe1}
\begin{eqnarray}
	0\!&\!=\!&\!\frac{f''}{f}+\frac{f'}{f}\bigg(\frac{1}{R}-\frac{h'}{h}+\frac{3}{2}\frac{\omega'}{\omega}\bigg)-\frac{5}{\ell^{2}}h^{2},\label{fe11}\\
	0\!&\!=\!&\!\bigg(\frac{f'}{f}+\frac{1}{R}+\frac{\omega'}{2\omega}-\frac{h'}{h}\bigg) \frac{\omega'}{\omega}+\frac{\omega''}{\omega}-h^2\bigg(\frac{10}{\ell^{2}}+\frac{4k}{\omega}\bigg),\nonumber\\
	\!&\!~\!&\!\label{fe12}\\
	0\!&\!=\!&\!\frac{f'}{f}-\frac{h'}{h}-\frac{5}{\ell^{2}}Rh^2+\frac{3}{2}\frac{\omega'}{\omega},\label{fe13}\\
	0\!&\!=\!&\!\frac{f''}{f}-\frac{5}{\ell^{2}}h^2+\frac{3}{2}\frac{\omega''}{\omega}-\bigg(\frac{f'}{f}+\frac{1}{R}+\frac{3}{2}\frac{\omega'}{\omega}\bigg)\frac{h'}{h}-\frac{3}{4}\frac{\omega'^2}{\omega^{2}},\nonumber\\
	\!&\!~\!&\!\label{fe14}
\end{eqnarray}
\end{subequations}
where primes denotes the derivative with respect to $R$. We have redefined the bulk cosmological constant as $\Lambda_{6}^{~}=-(D\!-\!2)(D\!-\!1)/(2\ell^{2}_{~})$ with $D=6$ denoting the dimension of the spacetime. The parameter $\ell$ is usually called the $\text{AdS}_{6}^{~}$ radius.

Here we should note that, in this paper, we do not consider the emergence of the Newtonian gravity on the brane. So if we obtain a Minkowski bulk spacetime with infinite volume from the above field equations~\eqref{fe1}, the four-dimensional gravity on the brane would deviate from GR at both large and small distances. To avoid this phenomenon, one could introduce the screening mechanism or the localization mechanism. The screening mechanism was proposed in the Dvali-Gabadadze-Porrati { (DGP)} model~\cite{Hernandez:2021qfp,Corman1,Abbott1,Pardo1,Deffayet1}. In this model, the bulk spacetime is also Minkowski. { So the gravity can propagate in the bulk freely. However, by introducing quantum corrections, the model has a four-dimensional Newtonian interaction on a zero-tension brane, which can be responsible for the Newtonian gravity at small distances in our four-dimensional universe. Unlike the DGP model, the RS-2 model~\cite{Randall2} introduces a non-zero brane tension. Due to the back-reaction of the tension, the bulk spacetime becomes AdS and then GR can be recovered at large distances in our four-dimensional universe by the localization mechanism}. However, one finds that both mechanisms do not affect the behavior of GWs in the bulk spacetime. In this paper, for the sake of simplification, we will not introduce these mechanisms in the model, and we will consider an $\text{AdS}_{6}^{~}$ spacetime. The corresponding solution could be found by assuming
\begin{equation}\label{so1}
	\omega(R)=R^{2},
\end{equation}
with which the field equations (\ref{fe1}) could be further simplified to
\begin{subequations}\label{fe2}
\begin{eqnarray}
	\frac{5}{\ell^{2}_{~}}h^{2}_{~}+\bigg(\frac{h'}{h}-\frac{4}{R}\bigg)\frac{f'}{f}-\frac{f''}{f}&\!=\!&0,\label{fe21}\\
	\frac{3}{R}-\bigg(\frac{2k}{R}+\frac{5R}{\ell^{2}_{~}}\bigg)h^{2}+\frac{f'}{f}-\frac{h'}{h}&\!=\!&0,\label{fe22}\\
	\frac{3}{R}-\frac{5R}{\ell^{2}_{~}}h^{2}+\frac{f'}{f}-\frac{h'}{h}&\!=\!&0,\label{fe23}\\
	\frac{5}{\ell^{2}}h^{2}+\bigg(\frac{f'}{f}+\frac{4}{R}\bigg)\frac{h'}{h}-\frac{f''}{f}&\!=\!&0.\label{fe24}
\end{eqnarray}	
\end{subequations}
Moreover, according to Eqs.~\eqref{fe22} and~\eqref{fe23}, $\omega(R)=R^{2}$ leads the maximally symmetric 3-space to be flat, i.e., $k=0$, which makes our following calculations much easier. Taking Eqs.~\eqref{fe21} and~\eqref{fe24} into account, one can obtain a relation between $f(R)$ and $h(R)$:
\begin{equation}\label{fe3}
	\frac{f'}{f}=-\frac{h'}{h}.
\end{equation}
Substituting it into Eq.~\eqref{fe23}, {we find that} the solution of $h(R)$ is
\begin{equation}\label{sh1}
	h^{2}_{~}(R)=\bigg[\Big(\frac{R}{\ell}\Big)^{2}_{~}-\Big(\frac{2\mathcal{M}}{R}\Big)^{3}_{~}\bigg]^{-1}_{~},
\end{equation}
where $\mathcal{M}$ is an integration constant denoting the effective mass of the gravitational configuration. Then, $f(R)$ is given by
\begin{equation}\label{sf1}
	f^{2}_{~}(R)=\Big(\frac{R}{\ell}\Big)^{2}_{~}-\Big(\frac{2\mathcal{M}}{R}\Big)^{3}_{~}.
\end{equation}
With the assumption $\omega(R)=R^{2}$ and the {solution} above, the metric (\ref{6metric1}) describing a six-dimensional AdS spacetime is rewritten as
\begin{equation}\label{6metric2} \text{d}s^{2}_{6}=-f^{2}_{~}(R)\text{d}T^{2}_{~}+R^{2}_{~}\text{d}\psi^{2}_{~}+f^{-2}_{~}(R)\text{d}R^{2}_{~}+R^{2}\text{d}\Sigma^{2}_{3},
\end{equation}
where
\begin{equation}
	\text{d}\Sigma^{2}_{3}=\text{d}r^{2}+r^{2}(\text{d}\theta^{2}+\text{sin}^{2}\theta \text{d}\phi^{2})
\end{equation}
represents a flat 3-subspace.

In the model, our homogeneous and isotropic universe is regarded as a codimension-2 brane in the background spacetime. Assuming that the 4-brane has a motion along the extra dimensions, its position in the bulk is then described by
\begin{subequations}\label{pa1}
  \begin{eqnarray}
	R&\!=\!&\mathcal{R}(\lambda),\label{pa11}\\
	\psi&\!=\!&\Psi(\lambda),\label{pa12}
  \end{eqnarray}
\end{subequations}
where $\lambda$ is an affine parameter. Then, the bulk time on the 4-brane can also be expressed in terms of the affine parameter as follows:
\begin{equation}\label{pa2}
	T=\mathcal{T}(\lambda).
\end{equation}
Substituting these parameterized variables~\eqref{pa1} and~\eqref{pa2} into the background metric~\eqref{6metric2}, we obtain the induced metric on the 4-brane,
\begin{equation}\label{4metric1} \text{d}s_{4}^{2}=-(f^{2}_{~}\dot{\mathcal{T}}^{2}_{~}-\mathcal{R}^{2}_{~}\dot{\Psi}^{2}_{~}-f^{-2}_{~}\dot{\mathcal{R}}^{2}_{~})\text{d}\lambda^{2}_{~}+\mathcal{R}^{2}_{~}\text{d}\Sigma^{2}_{3},
\end{equation}
where the dots denote the derivative with respect to the affine parameter $\lambda$. By setting $\lambda=t$, the induced metric~\eqref{4metric1} could coincide with the Friedmann-Lema\^{\i}tre-Robertson-Walker (FLRW) metric with $k=0$:
\begin{equation}\label{4metric2}
	\text{d}s_{4}^{2}=-\text{d}t^{2}_{~}+a^{2}_{~}(t)\text{d}\Sigma^{2}_{3}.
\end{equation}
So the six-dimensional AdS spacetime with the metric~\eqref{6metric2} could exactly describe a four-dimensional homogeneous and isotropic flat universe. The premise is that the universe is identified as a 4-brane with the induced metric~\eqref{4metric1} embedded in the bulk spacetime. Obliviously, such identification requires a connection between the bulk time $T$ and the cosmic time $t$ as follows:
\begin{equation}\label{tT1}
	\text{d}T^{2}_{~}=\frac{1+\mathcal{R}^{2}_{~}\dot{\Psi}^{2}_{~}+f^{-2}_{~}\dot{\mathcal{R}}^{2}_{~}}{f^{2}_{~}}\text{d}t^{2}_{~}.
\end{equation}
Moreover, comparing the metrics~\eqref{4metric1} and~\eqref{4metric2}, the 4-brane motion along the $R$ direction should be identified as the scale factor $a(t)$ of the FLRW metric:
\begin{equation}
	\mathcal{R}^{2}_{~}(t)=a^{2}_{~}(t),
\end{equation}
which means that the expansion of the universe is related to the 4-brane motion in the bulk.

In the braneworld theory, GWs are usually allowed to propagate in the bulk. Assuming that they have the speed of light, their trajectories should follow the six-dimensional null geodesics, which could deviate from the ``null geodesics'' defined by the induced metric on the brane, if the extrinsic curvature of the 4-brane is nonvanishing~\cite{Ishihara1}.
Therefore, for a given time interval on the 4-brane, the deviation might result in a discrepancy between the gravitational horizon radius (i.e., the projection of the horizon radius for the causal propagation of GWs {onto the 4-brane}) and the photon horizon radius (i.e., the horizon radius for the causal propagation of lights). Generally speaking, such a discrepancy depends on the bulk structure and 4-brane structure, so one can use it to investigate extra dimensions. In the next section, we will derive the gravitational horizon radius and the photon horizon radius in the model.

%From a phenomenological point of view, the deviation might result in a speed discrepancy between GWs and lights observed on the brane. Such a speed discrepancy depends on both of the bulk structure and 4-brane structure. For a given time interval on the 4-brane, one can measure the discrepancy by comparing gravitational horizon radius (i.e., the projection of the horizon radius for the causal propagation of GWs {onto the 4-brane}) with the photon horizon radius (i.e., the horizon radius for the causal propagation of lights). In the next section, we will derive the gravitational horizon radius and the photon horizon radius in the model.

\section{Horizon Radius}\label{sec3}

We calculate the gravitational horizon radius first. Assume that a GW signal is emitted by a source in our universe at the cosmic time $t_{A}^{~}$ and is detected by an observer at the cosmic time $t_{B}^{~}$. If it propagates at the light speed {in the bulk} with fixed $\theta$ and $\phi$, the corresponding gravitational horizon radius $r_{g}^{~}$ could be written as
\begin{equation}\label{ghr1}
	r_{g}^{~}=\int_{r_{A}^{~}}^{r_{B}^{~}}\text{d}r,
\end{equation}
where $r_{A}^{~}$ and $r_{B}^{~}$ are the radial coordinate distances of the source and the observer, respectively. In the four-dimensional GR, the gravitational horizon radius of a GW signal equals to the photon horizon radius of a light signal, if the two signals are simultaneously originated from the source and propagate for the same time interval in the universe. However, when extra dimensions exist, the GW can escape from the 4-brane and propagate in the bulk, which could make the gravitational horizon radius larger than the photon horizon radius. In the following, we will show how extra dimensions affect the gravitational horizon radius.

We label the source point of the GW as point $A$ and the detection point of the GW as $B$, both of which are fixed on the brane. Assuming the propagation speed of the GW to be the light speed, the corresponding trajectory is a six-dimensional null geodesic, which is given by the following equation:
\begin{equation}\label{6ng1}
    -f^{2}_{~}(R)\text{d}T^{2}_{~}+R^{2}_{~}\text{d}\psi^{2}_{~}+f^{-2}_{~}(R)\text{d}R^{2}_{~}+R^{2}\text{d}r^{2}_{~}=0.
\end{equation}
One can find three Killing vectors defined on the trajectory, $\mathcal{K}_{T}^{M}=(1,0,0,0,0,0)$, $\mathcal{K}_{r}^{M}=(0,0,0,1,0,0)$, and $\mathcal{K}_{\psi}^{M}=(0,1,0,0,0,0)$. Based on these Killing vectors, one can define three conserved quantities:
\begin{equation}\label{cq1}
	\kappa_{i}^{~}\equiv g_{KL}^{~}U^{K}_{~}\mathcal{K}^{L}_{i},
\end{equation}
where $i=T,r,\psi$ and $U^{M}_{~}=dx^{M}_{~}/d\lambda$ is a unit spacelike vector tangent to the geodesic. With the conserved quantities~\eqref{cq1} and Eq.~\eqref{6ng1}, the six-dimensional null
geodesic satisfies
\begin{subequations}\label{eom1}
	\begin{eqnarray} \bigg(\frac{\text{d}R}{\text{d}\lambda}\bigg)^{2}_{~}&\!=\!&\kappa^{2}_{T}-\frac{\kappa^{2}_{\psi}+\kappa^{2}_{r}}{R^{2}}f^{2}_{~},\label{eom11}\\
\bigg(\frac{\text{d}T}{\text{d}\lambda}\bigg)^{2}_{~}&\!=\!&\frac{\kappa^{2}_{T}}{f^{4}_{~}},\label{eom12}\\
\bigg(\frac{\text{d}r}{\text{d}\lambda}\bigg)^{2}_{~}&\!=\!&\frac{\kappa^{2}_{r}}{R^{4}_{~}},\label{eom13}\\
\bigg(\frac{\text{d}\psi}{\text{d}\lambda}\bigg)^{2}_{~}&\!=\!&\frac{\kappa^{2}_{\psi}}{R^{4}_{~}}.\label{eom14}
	\end{eqnarray}
\end{subequations}
One can find a useful relation between $R$ and $r$ by substituting Eq.~\eqref{eom11} into Eq.~\eqref{eom13}:
\begin{equation}\label{eom2} \text{d}R^{2}_{~}=\bigg[\frac{\kappa_{T}^{2}}{\kappa^{2}_{r}}R^{2}_{~}-\Big(\frac{\kappa^{2}_{\psi}}{\kappa^{2}_{r}}+1\Big)f^{2}_{~}\bigg]R^{2}_{~}\text{d}r^{2}_{~}.
\end{equation}
With the help of this relation, the gravitational horizon radius~\eqref{ghr1} can be expressed in terms of the coordinate locations of the two points on the $R$ direction, $R_{A}^{~}$ and $R_{B}^{~}$, as
\begin{equation}	r_{g}^{~}=\int_{R_{A}^{~}}^{R_{B}^{~}}\frac{1}{R}\bigg[\frac{\kappa_{T}^{2}}{\kappa^{2}_{r}}R^{2}_{~}-\Big(\frac{\kappa^{2}_{\psi}}{\kappa^{2}_{r}}+1\Big)f^{2}_{~}\bigg]^{-\frac{1}{2}}_{~}\text{d}R,
\end{equation}
from which, one can get the following relation:
\begin{equation}\label{ghr2}	\frac{1}{R_{A}^{~}}-\frac{1}{R_{B}^{~}}=\sqrt{\frac{\kappa_{T}^{2}}{\kappa^{2}_{r}}\ell^{2}_{~}-\frac{\kappa^{2}_{\psi}}{\kappa^{2}_{r}}-1}\,\frac{r_{g}^{~}}{\ell}.
\end{equation}
Note that, to get the above relation we have ignored the contribution of the effective mass $\mathcal{M}$. Here and after, we will set $\mathcal{M}=0$ for simplicity.

To eliminate the constants of motion in Eq.~\eqref{ghr2}, we recall the six-dimensional null geodesic equations. Substituting Eq.~\eqref{eom12} into Eq.~\eqref{eom13}, we have the following relation between $T$ and $r$:
\begin{equation}
	\text{d}T_{~}=\frac{\kappa_{T}}{\kappa_{r}}\ell^{2}_{~}\text{d}r_{~},
\end{equation}
with which the gravitational horizon radius could be expressed in terms of the bulk time interval as
\begin{equation}\label{ghr3}
	T_{B}^{~}-T_{A}^{~}=\frac{\kappa_{T}^{~}}{\kappa_{r}^{~}}\ell^{2}_{~}r_{g}^{~},
\end{equation}
where $T_{A}^{~}$ and $T_{B}^{~}$ correspond to the emission time and detection  time of the GW, respectively. Similarly, $r$ is related to $\psi$ through the combination of Eqs.~\eqref{eom13} and~\eqref{eom14}:
\begin{equation}
	\text{d}\psi_{~}=\frac{\kappa_{\psi}}{\kappa_{r}}\text{d}r_{~}.
\end{equation}
Then, one can express the gravitational horizon radius as
\begin{equation}\label{ghr4}
	\psi_{B}^{~}-\psi_{A}^{~}=\frac{\kappa_{\psi}^{~}}{\kappa_{r}^{~}}r_{g}^{~},
\end{equation}
where $\psi_{A}^{~}$ and $\psi_{B}^{~}$ are values of the polar angular of points $A$ and $B$, respectively. Taking advantage of Eqs.~\eqref{ghr2},~\eqref{ghr3}, and~\eqref{ghr4}, one finally finds an expression of the gravitational horizon radius without the constants of motion:
\begin{equation}\label{ghr4x}	r_{g}^{2}=\bigg(\frac{T_{B}^{~}-T_{A}^{~}}{\ell}\bigg)^{2}_{~}-\ell^{2}_{~}\bigg(\frac{1}{R_{A}^{~}}-\frac{1}{R_{B}^{~}}\bigg)^{2}_{~}-(\psi_{B}^{~}-\psi_{A}^{~})^{2}_{~}.
\end{equation}
On the other hand, since the source and the observer are both located on the brane, one can convert all the quantities in Eq.~\eqref{ghr4x} into observable quantities.

Recalling the relation~\eqref{tT1}, the bulk time interval can be rewritten as
\begin{equation}	T_{B}^{~}-T_{A}^{~}=\int_{t_{A}^{~}}^{t_{B}^{~}}\frac{\sqrt{1+f^{-2}_{~}\dot{\mathcal{R}}^{2}_{~}+\mathcal{R}^{2}_{~}\dot{\Psi}^{2}_{~}}}{f}\,\text{d}t.
\end{equation}
Together with the relation~\eqref{ghr3}, it gives
\begin{equation}\label{tT2}	\frac{\kappa_{T}^{2}}{\kappa_{r}^{2}}\ell^{2}_{~}=\frac{1}{r_{g}^{2}}\bigg(\int_{t_{A}^{~}}^{t_{B}^{~}}\sqrt{\frac{1}{\mathcal{R}^{2}_{~}}+\frac{\ell^{2}_{~}}{\mathcal{R}^{2}_{~}}H^{2}_{~}+\dot{\Psi}^{2}_{~}}\,\text{d}t\bigg)^{2}_{~},
\end{equation}
where $H=\mathcal{\dot R}/\mathcal{R}$ is the Hubble parameter. In this paper, we neglect the back-reaction of the 4-brane on the bulk spacetime, so it is convenient to suppose the 4-brane as a test particle. Then, employing the standard procedure usually used to investigate a test particle around a black hole, one can obtain the following 4-brane motion on the $\Psi$ direction:
\begin{equation}
	\frac{\partial\Psi}{\partial T}=\frac{L}{E}\frac{1}{\ell^{2}_{~}},
\end{equation}
where $E$ and $L$ are respectively the energy and polar angular momentum of the brane. With the relation~\eqref{tT1}, this expression turns into {
\begin{equation}
	\dot{\Psi}=\frac{1}{\mathcal{R}}\sqrt{\frac{L^{2}_{~}}{E^{2}_{~}\ell^{2}_{~}-L^{2}_{~}}}\sqrt{1+\ell^{2}_{~}H^{2}_{~}}.
\end{equation}}
In the following, we will use $\dot{\Psi}=C/\mathcal{R}$ for simplicity with $C$ being a parameter relating to $E$, $L$, and $\ell$.
%And we will set its value as $0\leqslant C\leqslant 1$.
Then, the right-hand side of Eq.~\eqref{tT2} can be expanded under the low-energy limit $\ell H\ll 1$ as
\begin{eqnarray}\label{tT3}
	\frac{\kappa_{T}^{2}}{\kappa_{r}^{2}}\ell^{2}_{~}
	&\!\approx\!&
	\frac{1}{r_{g}^{2}}\bigg\{\int_{t_{A}^{~}}^{t_{B}^{~}}\frac{1}{\mathcal{R}}
	\Big[\sqrt{1+C^{2}_{~}}+\frac{H^{2}_{~}\ell^{2}_{~}}{2\sqrt{1+C^{2}_{~}}}\nonumber\\
	&\!~\!&	-\frac{H^{4}_{~}\ell^{4}_{~}}{8(1+C^{2}_{~})^{3/2}_{~}}+\mathcal{O}(H^{6}_{~}\ell^{6}_{~})\Big]\text{d}t
	\bigg\}^{2}_{~}.
\end{eqnarray}
In addition, since $\dot{\Psi}=C/\mathcal{R}$, the polar angular interval between the emission and detection of the GW can be given by
\begin{equation}
	\psi_{B}^{~}-\psi_{A}^{~}=\int_{t_{A}^{~}}^{t_{B}^{~}}\frac{C}{\mathcal{R}}\text{d}t,
\end{equation}
which combined with Eq.~\eqref{ghr4} yields
\begin{equation}\label{tpsi1}	\frac{\kappa_{\psi}^{2}}{\kappa_{r}^{2}}=\frac{1}{r_{g}^{2}}\bigg(\int_{t_{A}^{~}}^{t_{B}^{~}}\frac{C}{\mathcal{R}}\text{d}t\bigg)^{2}_{~}.
\end{equation}
Note that the left-hand side of Eq.~\eqref{ghr2} can be  converted into an integral form:
\begin{equation}\label{tR1}
\frac{1}{R_{A}^{~}}-\frac{1}{R_{B}^{~}}=\int_{t_{A}^{~}}^{t_{B}^{~}}\frac{H}{\mathcal{R}}\text{d}t=	\sqrt{\frac{\kappa_{T}^{2}}{\kappa^{2}_{r}}\ell^{2}_{~}-\frac{\kappa^{2}_{\psi}}{\kappa^{2}_{r}}-1}\,\frac{r_{g}^{~}}{\ell}.
\end{equation}
Substituting Eqs.~\eqref{tT3} and~\eqref{tpsi1} into Eq.~\eqref{tR1} to eliminate the constants of motion, the gravitational horizon radius reads
\begin{strip}
\begin{equation}\label{ghr42}
	r_{g}^{2}=
	\bigg\{\int_{t_{A}^{~}}^{t_{B}^{~}}\frac{1}{\mathcal{R}}	\Big[\sqrt{1+C^{2}_{~}}+\frac{H^{2}_{~}\ell^{2}_{~}}{2\sqrt{1+C^{2}_{~}}}-\frac{H^{4}_{~}\ell^{4}_{~}}{8(1+C^{2}_{~})^{3/2}_{~}}
	+\mathcal{O}(H^{6}_{~}\ell^{6}_{~})\Big]\text{d}t
	\bigg\}^{2}_{~}
	-\bigg(\int_{t_{A}^{~}}^{t_{B}^{~}}\frac{C}{\mathcal{R}}\text{d}t\bigg)^{2}_{~}
	-\ell^{2}_{~}\bigg(\int_{t_{A}^{~}}^{t_{B}^{~}}\frac{H}{\mathcal{R}}\text{d}t\bigg)^{2}_{~}.
\end{equation}
\end{strip}
Then we express the radius in terms of observables. In braneworld models, the evolution of the universe is influenced by extra dimensions, so the corresponding brane cosmology might deviate from the conventional cosmology. However, the correction of extra dimensions to cosmology usually appear in the higher-order terms. For example, in the RS-1 and RS-2 models the Hubble parameter follows~\cite{Csaki1,Cline1}
\begin{equation}
	H^2\propto (\rho+c_0^{~}\rho^2),
\end{equation}
where $\rho$ is the energy density of the brane matter and $c_0^{~}$ is a model-dependent parameter. It is obvious that the brane cosmology in the models gives the conventional cosmology as $H^{2}_{~}\propto \rho$ in the leading order. And the correction to the $\Lambda$CDM model is at the order $\rho^2$, which is important when the temperature of the universe is higher than $1\,\text{TeV}$~\cite{Cline1}, i.e., the correction term dominates only in the early stage of the universe. It means that the departure between brane cosmology and the $\Lambda$CDM model could be neglected for the late universe. In this paper, since we only study GW events occurring in the late universe, we do not consider the correction from extra dimensions to the brane cosmology for convenience.

Taking advantage of it, we will use the $\Lambda$CDM model to describe our four-dimensional universe. On the other hand, since our research does not involve the early stage of the universe, we neglect the radiation in the universe. { Following the procedure in Ref.~\cite{Visinelli1}, we then set the redshift at point $B$ as $z_{B}^{~}=0$.} Further, by converting the integrals over the cosmic time into the integrals over the redshift $z$, we can express the gravitational horizon radius as
\begin{strip}
\begin{eqnarray}\label{ghr5}
	\tilde{r}_{g}^{2}
	&\!\approx\!&	\bigg[\int_{0}^{z_{A}^{~}}\frac{1}{H_{B}^{~}\sqrt{\Omega_{\Lambda}^{~}+\Omega_{m}^{~}(1+z)^{3}_{~}}}\text{d}z\bigg]^{2}_{~}
+\ell^{2}_{~}\bigg[\int_{0}^{z_{A}^{~}}\frac{1}{\sqrt{\Omega_{\Lambda}^{~}+\Omega_{m}^{~}(1+z)^{3}_{~}}}\text{d}z\bigg]
	\bigg[\int_{0}^{z_{A}^{~}}\sqrt{\Omega_{\Lambda}^{~}+\Omega_{m}^{~}(1+z)^{3}_{~}}\,\text{d}z\bigg]
	\nonumber\\
	&\!~\!&	-\frac{H_{B}^{2}\ell^{4}_{~}}{4(1+C^{2}_{~})}\bigg[\int_{0}^{z_{A}^{~}}\frac{1}{\sqrt{\Omega_{\Lambda}^{~}+\Omega_{m}^{~}(1+z)^{3}_{~}}}\text{d}z\bigg]
\bigg\{\int_{0}^{z_{A}^{~}}\Big[\Omega_{\Lambda}^{~}+\Omega_{m}^{~}(1+z)^{3}_{~}\Big]^{3/2}_{~}\,\text{d}z\bigg\}
	\nonumber\\
	&\!~\!&	+\frac{H_{B}^{2}\ell^{4}_{~}}{4(1+C^{2}_{~})}\bigg[\int_{0}^{z_{A}^{~}}\sqrt{\Omega_{\Lambda}^{~}+\Omega_{m}^{~}(1+z)^{3}_{~}}\,\text{d}z\bigg]^{2}_{~}
	-\ell^{2}_{~}\left(\int_{0}^{z_{A}^{~}}\text{d}z\right)^{2}_{~}+\mathcal{O}(H^{4}_{B}\ell^{6}_{~}),
\end{eqnarray}
\end{strip}
where $H_{B}^{~}$ is the Hubble parameter at cosmic time $t_{B}^{~}$ and $z_{A}^{~}$ is the source redshift measured by the GW observation. Both the density parameters $\Omega_{\Lambda}^{~}$ (dark energy) and $\Omega_{m}^{~}$ (nonrelativistic matter) take the values at cosmic time $t_{B}^{~}$. Note that $r_{g}^{~}$ is already rescaled by setting $\tilde{r}_{g}^{~}=\mathcal{R}_{B}^{~}r_{g}^{~}$ with $\mathcal{R}_{B}^{~}$ the scale factor at the cosmic time $t_{B}^{~}$. The scale factor is correspondingly rescaled by $\tilde{\mathcal{R}}=\mathcal{R}/\mathcal{R}_{B}^{~}$, so that the value of the rescaled scale factor at $t_{B}^{~}$ is unit. Note that the expression~\eqref{ghr5} could recover the result obtained in the five-dimensional AdS model~\cite{Visinelli1}, when the 4-brane does not move on the $\psi$ direction, i.e., $C=0$. Moreover, the motion of the 4-brane on the $\psi$ direction does not contribute to the gravitational horizon radius until up to the order $H^{2}_{B}\ell^{4}$.
%Therefore, the deviation from the five-dimensional case is slight under the low-energy approximation.
%And it is likely to give a observable contribution to the horizon radius as the redshift of the source increases. Given that, we will reserve the terms $\sim H^{2}_{B}\ell^{4}_{~}$ in Eq.~\eqref{ghr5}.
%Note that in this paper, we concern not only the GWs from a source with a low redshift, but also the GWs from a high-redshift source. Especially, we will first put the emission of GW170817 to the point $A$ as an example, and then consider a more general case that the source redshift becomes larger. Accordingly, we do not use the usual approach, i.e., the low-redshift approximation ($z\ll1$), to give the result of each of integrals on the right-hand side of the expression~\eqref{ghr5}.
So its contribution is negligible when $z_{A}^{~}\ll1$ and $H^{2}_{B}\ell^{4}\ll1$. However, as the redshift of the source increases, those terms are likely to dominate the gravitational horizon radius. Integrating the right-hand side of Eq.~\eqref{ghr5}, the gravitational horizon radius is finally given by
%Then, to integrate those terms on the right-hand-side of Eq.~\eqref{ghr5}, we would like to introduce the parametrization $X\equiv(1+z)^{3}\Omega_{m}^{~}/\Omega_{\Lambda}^{~}$, with which the horizon radius finally turns into
\begin{eqnarray}\label{ghr6}
	\tilde{r}_{g}^{2}&\!\approx\!&\frac{1}{H_{B}^{2}}\bigg[
	\frac{1}{\Omega_{\Lambda}^{~}}W_{1}^{2}+\ell^{2}_{~}H_{B}^{2}(W_{1}^{~}W_{2}^{~}-z_{A}^{2})\nonumber\\
	&\!~\!&
	+\frac{\ell^{4}_{~}H_{B}^{4}\Omega_{\Lambda}^{~}}{4(1+C^{2}_{~})}(W^{2}_{2}-W^{~}_{1}W^{~}_{3})\bigg],
\end{eqnarray}
where we reserve the terms up to the order $H_{B}^{2}\ell^{4}_{~}$. Here, $W_{1}^{~}$, $W_{2}^{~}$, and $W_{3}^{~}$ are parameter functions defined by
\begin{subequations}
    \begin{eqnarray}	    W_{1}^{~}&\!\equiv\!&(1+z_{A}^{~})~^{~}_{2}F_{1}^{~}\!\Big[\frac{1}{3},\frac{1}{2};\frac{4}{3};-\frac{\Omega_{m}^{~}}{\Omega_{\Lambda}^{~}}(1+z_{A}^{~})^{3}_{~}\Big]\nonumber\\
&\!~\!&-~^{~}_{2}F_{1}^{~}\!\Big(\frac{1}{3},\frac{1}{2};\frac{4}{3};-\frac{\Omega_{m}^{~}}{\Omega_{\Lambda}^{~}}\Big),\\
W_{2}^{~}&\!\equiv\!&(1+z_{A}^{~})~^{~}_{2}F_{1}^{~}\!\Big[\!\!-\!\frac{1}{2},\frac{1}{3};\frac{4}{3};-\frac{\Omega_{m}^{~}}{\Omega_{\Lambda}^{~}}(1+z_{A}^{~})^{3}_{~}\Big]\nonumber\\
&\!~\!&-~^{~}_{2}F_{1}^{~}\!\Big(\!\!-\!\frac{1}{2},\frac{1}{3};\frac{4}{3};-\frac{\Omega_{m}^{~}}{\Omega_{\Lambda}^{~}}\Big),\\
W_{3}^{~}&\!\equiv\!&(1+z_{A}^{~})~^{~}_{2}F_{1}^{~}\!\Big[\!\!-\!\frac{3}{2},\frac{1}{3};\frac{4}{3};-\frac{\Omega_{m}^{~}}{\Omega_{\Lambda}^{~}}(1+z_{A}^{~})^{3}_{~}\Big]\nonumber\\
&\!~\!&-~^{~}_{2}F_{1}^{~}\!\Big(\!\!-\!\frac{3}{2},\frac{1}{3};\frac{4}{3};-\frac{\Omega_{m}^{~}}{\Omega_{\Lambda}^{~}}\Big),
    \end{eqnarray}
\end{subequations}
respectively. Here $^{~}_{2}F^{~}_{1}\!(a,b;c;d)$ is just the Gaussian hypergeometric function.

Now, let us calculate the photon horizon radius on the brane. In the braneworld model, particles in the Standard Model of particle physics are all confined on the 4-brane. Thus the trajectory of a photon is just a four-dimensional ``null geodesic'' on the brane. Recalling the induced metric~\eqref{4metric2} (the FLRW metric), the four-dimensional ``null geodesic'' with $d\theta=0$ and $d\phi=0$ follows
\begin{equation}\label{4metric3}
	-\text{d}t^{2}_{~}+\mathcal{R}^{2}_{~}(t)\text{d}r^{2}_{~}=0.
\end{equation}
Assume that an EMW signal is emitted from the point $A$ and detected at the point $C$. The horizon radius of the EMW signal on the $r$ direction during the cosmic time interval $t_{C}^{~}-t_{A}^{~}$ is then given by
\begin{equation}\label{xxx}	
    r_{\gamma}^{~}=\int_{r_{A}^{~}}^{r_{C}^{~}}\text{d}r=
    \int_{t_{A}^{~}}^{t_{C}^{~}}\frac{1}{\mathcal{R}}\text{d}t,
\end{equation}
where $r_{A}^{~}$ and $r_{C}^{~}$ are the radial coordinate distances of the EMW signal at the cosmic times $t_{A}^{~}$ and $t_{C}^{~}$ on the 4-brane, respectively.

\section{Time Delay}\label{sec4}

We now consider a GW signal and an EMW signal simultaneously originated with the light speed from the same source point $A$. Putting the source at the origin of the coordinate system $(r,\theta,\phi)$, the trajectories of the two signals have $d\theta=0$ and $d\phi=0$. If the signals are finally detected by the same observer at $r_{B}^{~}$, the existence of extra dimensions could cause a difference between their trajectories. Consequently, the two signals will reach the observer successively, i.e., there is a time delay between the detections of these two signals. We can set $t_{B}^{~}$ as the cosmic time when the GW signal reaches the observer, and $t_{C}^{~}$ as the moment when the EMW signal reaches the observer. %Before going ahead, we shall rewrite the photon horizon radius in terms of observables as follows:
To facilitate the comparison of the gravitational horizon radius and the photon horizon radius, we give the photon horizon radius during the time interval $t_{B}^{~}-t_{A}^{~}$ as
\begin{equation}\label{phr1}
	\tilde{r}_{\gamma}= \mathcal{R}_{B}^{~}r_{\gamma}^{~}
= \mathcal{R}_{B}^{~}\int_{t_{A}^{~}}^{t_{B}^{~}}\frac{1}{\mathcal{R}}\text{d}t
=\frac{W_{1}}{H_{B} \sqrt{\Omega_{\Lambda}}} ,
\end{equation}
where we have used $\tilde{r}_{\gamma}^{~}=\mathcal{R}_{B}^{~}r_{\gamma}^{~}$ to rescale $r_{\gamma}^{~}$. It is found that the photon horizon radius~\eqref{phr1} just equals the leading-order term of the gravitational horizon radius~ (\ref{ghr6}). It means that all the high-order terms in the expression~\eqref{ghr6} come from the contribution of the extra dimensions, and that the GW signal will arrive at the observer before the EMW signal in the model. Therefore, the time delay $\Delta t$ can be defined as the time interval between $t_{C}^{~}$ and $t_{B}^{~}$, i.e., $\Delta t\equiv t_{C}^{~}-t_{B}^{~}$.

Since the comoving distances from the source to the observer for the GW and EMW signals are the same, one should have
\begin{equation}\label{ggd1}	\tilde{r}_{g}=\int_{t_{A}^{~}}^{t_{C}^{~}}\frac{1}{\tilde{\mathcal{R}}}\text{d}t
= \tilde{r}_{\gamma}^{~}+\int_{t_{B}^{~}}^{t_{B}^{~}+\Delta t}\frac{1}{\tilde{\mathcal{R}}}\text{d}t{~},
\end{equation}
where the magnitude of the second term on the right-hand side reveals how effective the shortcut effect is. Obviously, the time delay increases with the source redshift. When the source redshift is small enough, for example the event GW170817 and its counterpart GRB 170817A, the change in the scale factor could be ignored during $\Delta t$. Then, the relation~\eqref{ggd1} can be well approximated as
\begin{equation}\label{ggd2}
	\tilde{r}_{g} \approx  \tilde{r}_{\gamma}^{~}+\Delta t  {~},
\end{equation}
which is also the normal practice in the previous researches~\cite{Yu1,Visinelli1,Lin1} under the low-redshift approximation. For a high-redshift source, the discrepancy between the trajectories of the GW  and EMW signals could be prominent (see Eqs.~\eqref{ghr6} and~\eqref{phr1}). Therefore, the time delay $\Delta t$ might be longer. In this case, the expansion of the universe during $\Delta t$ becomes nonnegligible and the approximation~\eqref{ggd2} is no longer available. So we should rescale the scale factor and the photon horizon radius by setting $\bar{\mathcal{R}}\equiv\mathcal{R}/\mathcal{R}_{C}^{~}$ and $\bar{r}\equiv r\mathcal{R}_{C}^{~}$ for the EMW observation. Here $\mathcal{R}_{C}^{~}$ is the scale factor at the cosmic time $t_{C}^{~}$. For a high-redshift source, the relation~\eqref{ggd1} turns to
\begin{eqnarray}
	\tilde{r}_{g}
	&\!=\!&\tilde{r}_{\gamma}^{~}	+\frac{\mathcal{R}_{B}^{~}}{\mathcal{R}_{C}^{~}}\int_{\bar{\mathcal{R}}_{B}^{~}}^{1}\!\!\frac{1}{H\bar{\mathcal{R}}^{2}_{~}}\text{d}\bar{\mathcal{R}}
	= \tilde{r}_{\gamma}^{~}	+\frac{\mathcal{R}_{B}^{~}}{\mathcal{R}_{C}^{~}H_{B}^{~}\sqrt{\Omega_{\Lambda}^{~}}}W_{4}^{~},\nonumber\\
	&\!~\!& ~
\end{eqnarray}
where $W_{4}^{~}$ is a parameter function. If we set
\begin{equation} 1+z_{A}^{~}=\frac{\mathcal{R}_{B}^{~}}{\mathcal{R}_{A}^{~}}~~~~\text{and}~~~~1+z'_{A}=\frac{\mathcal{R}_{C}^{~}}{\mathcal{R}_{A}^{~}},
\end{equation}
then one has
\begin{equation}\label{ggd3}
	\tilde{r}_{g}
	=\tilde{r}_{\gamma}^{~}
	+\frac{1+z_{A}^{~}}{1+z'_{A}}\frac{W_{4}}{H_{B}^{~}\sqrt{\Omega_{\Lambda}^{~}}}  {~},
\end{equation}
where
\begin{eqnarray}	
W_{4}^{~}&\!\equiv\!&\frac{1+z'_{A}}{1+z_{A}^{~}}\,_{2}^{~}F_{1}^{~}\!\Big[\frac{1}{3},\frac{1}{2};\frac{4}{3};-\frac{\Omega_{m}^{~}}{\Omega_{\Lambda}^{~}}\Big(\frac{1+z'_{A}}{1+z_{A}^{~}}\Big)^{3}_{~}\Big]\nonumber\\
&\!~\!&-~_{2}^{~}F_{1}^{~}\!\Big(\frac{1}{3},\frac{1}{2};\frac{4}{3};-\frac{\Omega_{m}^{~}}{\Omega_{\Lambda}^{~}}\Big).
\end{eqnarray}
Note that $z'_{A}$ is the source redshift measured by the EMW observation at the cosmic time $t_{C}^{~}$. As we can see from the relation~\eqref{ggd3}, the time delay between the two signals is converted into the difference between the source redshifts measured by the GW observation and EMW observation. And one can check that the gravitational horizon radius equals the photon horizon radius when $z'_{A}=z_{A}^{~}$.

\section{Results}\label{sec5}

In the previous sections, we obtain the expression of the gravitational horizon radius, and find that it equals to the photon horizon radius at the leading order. These two horizon radii can be connected through the identity~\eqref{ggd1}, which also connects the $\text{AdS}_{6}^{~}$ radius and the observable quantities on the 4-brane. Therefore, with the identity~\eqref{ggd1}, one can investigate the structure of the six-dimensional AdS spacetime on the 4-brane. Generally, the practice requires that the GW and EMW signals detected by the same observer are originated from the same source, and that the time interval between the emissions of the two signals is predictable. In astrophysics, a binary neutron star (BNS) merger is expected to be the source of both the GW and EMW signals. The target EMW signal from the BNS is a short gamma-ray burst (sGRB), most energy of which is however collimated into a narrow jet. Therefore, detecting the EM counterpart of an identified GW event requires the observer to be right within its narrow jet, which makes it rare to detect both the sGRB and GW from the same BNS. Fortunately, in 2017, the joint GW and EMW observations found the event GW170817 and a subsequent short gamma-ray burst (GRB 170817A)~\cite{Coulter1,Goldstein1,Savchenko1,Abbott2}. The analysis on the sky location of the host galaxy of GRB 170817A indicates that the two signals are originated from the same source---the coalescence of a BNS in NGC 4993~\cite{Coulter1,Pan1}. In the light of the joint observations, the EM counterpart GRB 170817A arrived at the earth $1.74_{-0.05}^{+0.05}\,$s later than GW170817~\cite{Coulter1,Goldstein1,Savchenko1,Abbott4,Pan1,Abbott3,Abbott2}. Such a time delay between GW170817 and GRB 170817A might support the existence of extra dimensions. And it could provide new constraints on various extra dimensional models~\cite{Pardo1,Abbott1,Hernandez:2021qfp,Visinelli1,Yu2,Lin1} as well. Next, we will use the event GW170817 and its EM counterpart GRB 170817A to constrain the $\text{AdS}_{6}^{~}$ radius in the model.

\subsection{Constraint for a low-redshift source}

For the event GW170817, we consider the source redshift as $z_{A}^{~}=0.008^{+0.002}_{-0.003}$, which is the result reported by the Laser Interferometer Gravitational-Wave Observatory (LIGO) and Virgo collaborations~\cite{Abbott2}. For simplicity, we can set the time lag between the detections of GW170817 and GRB 170817A as $1.74$s without the error. On the basis of Refs.~\cite{Shibata1,Rezzolla1,Paschalidis1,Rezzolla2,Ciolfi1,Tsang1}, we know that the sGRB and GW in the event may not be emitted at the same time and astrophysical models allow a time lag $(-100\,\text{s},1000\,\text{s})$ between the emissions of the sGRB and GW, which means that the emission of the sGRB could be $100\,\text{s}$ earlier or $1000\,\text{s}$ later than the emission of the GW. Since the gravitational horizon radius is larger than the photon horizon radius in our model, the reasonable time lag between the emissions of the sGRB and GW should be $(-100\,\text{s},1.74\,\text{s})$. Therefore, for the event GW170817, if we assume that the emissions of the sGRB and GW are simultaneous, the real time delay between the detections of the two signals should be $(0\,\text{s}, 101.74\,\text{s})$. It means that in the most extreme case, the GW signal can arrive $101.74\,\text{s}$ earlier than the sGRB signal. Obviously, the expansion of the universe during the time delay is negligible, so the low-redshift approximation~\eqref{ggd2} is accurate enough for studying GW170817/GRB 170817A in the model. { Recalling expressions~\eqref{ghr6} and~\eqref{phr1}, the relations among the $\text{AdS}_{6}^{~}$ radius, time delay, and source redshift, under the low-energy approximation, can be further expressed as
\begin{equation}
	\Delta t^{2}_{~} \!\approx\!\!
	\frac{1}{H_{B}^{2}}\!\bigg[
	\ell^{2}_{~}H_{B}^{2}(W_{1}^{~}\!W_{2}^{~}\!-\!z_{A}^{2})
	\!+\!\frac{\ell^{4}_{~}H_{B}^{4}\Omega_{\Lambda}^{~}}{4(1\!+\!C^{2}_{~})}(W^{2}_{2}\!-\!W^{~}_{1}\!W^{~}_{3})\bigg].
\end{equation}
For a given time delay, it gives the relation between the $\text{AdS}_{6}^{~}$ radius and the source redshift.} In the following, we set $H_{B}^{~}=67.66\,\text{km}\,\text{s}^{-1}_{~}\text{Mpc}^{-1}_{~}$, $\Omega_{\Lambda}^{~}=0.6889$, and $\Omega_{m}^{~}=0.3111$ based on the 2018 release of Planck satellite data~\cite{Planck1}.
\begin{figure}[!htb]
\center{
\subfigure[]{\includegraphics[width=3.8cm]{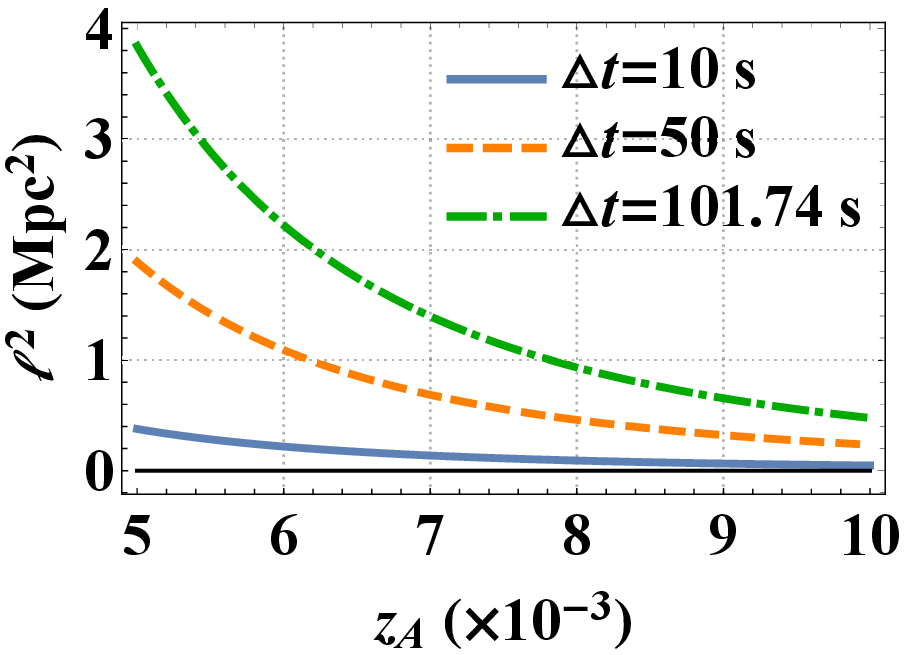}\label{constraint11}}
\subfigure[]{\includegraphics[width=4.2cm]{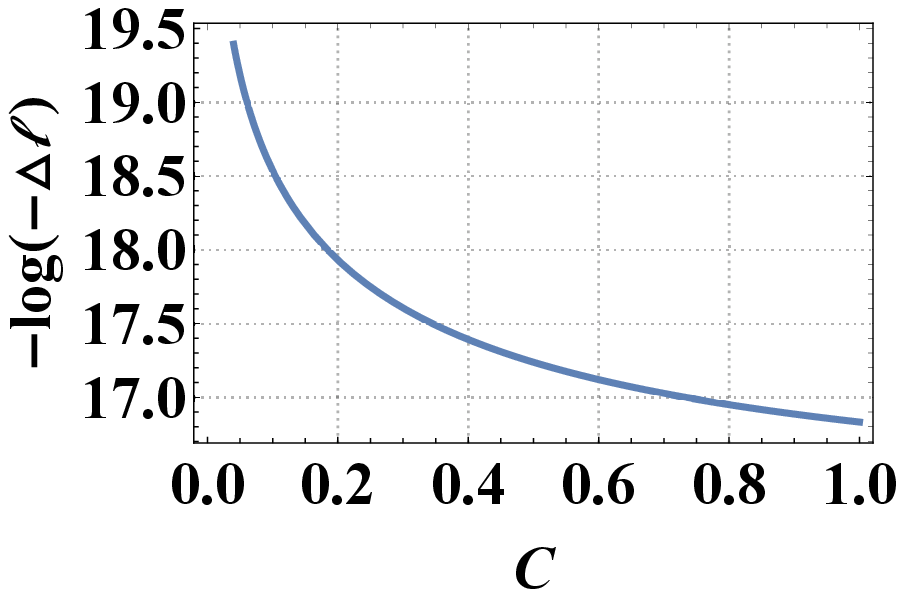}\label{constraint12}}
}
\caption{The constraint on the $\text{AdS}_{6}$ radius from the joint observations of GW170817/GRB 170817A. (a) The $\text{AdS}_{6}$ radius with respective to the source redshift. The redshift ranges from $z_{A}^{~}=0.005$ to $z_{A}^{~}=0.01$. The parameter $C$ is set to $C=1/2$. The values of the time delay between the detections of the two signals are chosen as $\Delta t=10\,\text{s}$ (blue solid curve), $\Delta t=50\,\text{s}$ (orange dashed curve), and $\Delta t=101.74\,\text{s}$ (green dashed and dotted curve). The black solid curve with $\Delta t=0\,\text{s}$ gives a lower boundary of the $\text{AdS}_{6}$ radius. (b) The contribution of the brane motion $\Psi(t)$ to the constraint on the $\text{AdS}_{6}^{~}$ radius. The values for the other parameters are $z_{A}^{~}=0.5$ and $\Delta t=101.74\,\text{s}$.}
\label{constraint1}
\end{figure}

In Fig.~\ref{constraint1}, we show the constraint on the $\text{AdS}_{6}^{~}$ radius with the joint observations of GW170817/GRB 170817A. From Fig.~\ref{constraint11}, the $\text{AdS}_{6}^{~}$ radius decreases as the source redshift increases for a given time delay. For a given source redshift, the $\text{AdS}_{6}^{~}$ radius increases with the time delay. We introduce a dimensionless quantity,
\begin{equation}
	\Delta\ell\equiv\frac{\ell-\ell_{0}^{~}}{\ell_{0}^{~}},
\end{equation}
to measure the contribution of the parameter $C$ to the constraint on the $\text{AdS}_{6}^{~}$ radius. Here $\ell_{0}^{~}$ is just the $\text{AdS}_{6}^{~}$ radius with $C=0$.
As $C$ increases, if $\Delta\ell$ changes a lot, then, $C$ is an important parameter for the constraint on the $\text{AdS}_{6}^{~}$ radius. From Fig.~\ref{constraint12}, one finds that $\Delta\ell$ is tiny for $0\leq C\leq1$. It means that the brane's motion $\Psi(t)$ contributes little to the gravitational horizon radius in the low-redshift case, which is consistent with our analysis in Sec.~\ref{sec3}. It is also found that $\Delta\ell$ is negative for $C>0$ and becomes smaller when $C$ goes larger. Therefore, when $C=0$, $z_{A}^{~}=0.005$, and $\Delta t=101.74\,\text{s}$, there exists an upper limit to the $\text{AdS}_{6}^{~}$ radius, i.e., $\ell^{2}_{~} \lesssim  3.84\,\text{Mpc}^{2}_{~}$.

\subsection{Prediction for a high-redshift source}

In the previous section, we obtain a constraint on the $\text{AdS}_{6}^{~}$ radius through the joint observations of GW170817/GRB 170817A. However, as we have emphasized, the relation~\eqref{ggd2} used there is valid only for the event with a short time delay, during which the cosmological expansion is negligible. Recently, LIGO and Virgo collaboration reported $35$ compact binary coalescence candidates identified up to the end of the second half of their third observing run~\cite{LIGOScientific1}. The redshifts of these candidates are all beyond $z=0.04$ and over 30 candidates have a redshift at the order $z\sim0.1$. There are even six candidates (such as GW200220\_061928, GW200308\_173609, and GW200322\_091133) whose redshifts are beyond $z=1$ (see Table~IV in Ref.~\cite{LIGOScientific1}). Therefore, it is expected that, in the future more and more GW events with the source redshifts $z>1$ will be detected by the next generation of ground-based GW detectors and the space-based GW detectors~\cite{Broeck1,Punturo1,Danzmann1,Seoane1,Luo1,Hu1,Seto1,Kawamura1,Kawamura2,Kramer1,Jenet1,Hobbs1,Hobbs2,Hobbs3,Gong1}. However, when the source redshift of a GW event goes larger, the time delay could be so long that the cosmological expansion is nonnegligible. In this case, the relation~\eqref{ggd2} under the low-redshift approximation is no longer applicable, so we have to use a more general relation~\eqref{ggd3} to study the shortcut of GWs. We use $z_{A}^{~}$ and $z'_{A}$ to denote the different source redshifts independently measured by the GW and EMW observations, respectively. With our previous study on the event GW170817/GRB 170817A, we set the $\text{AdS}_{6}^{~}$ radius as $\ell^{2}_{~}\approx 3.84\,\text{Mpc}^{2}_{~}$ in the future observations for simplicity. Moreover, we can define a dimensionless parameter to denote the relative deviation of the source redshift $z'_{A}$ with respective to the source redshift $z_{A}^{~}$:
\begin{equation}
	\Delta z\equiv\frac{z'_{A}-z_{A}^{~}}{z_{A}^{~}}.
\end{equation}
Then the relation~\eqref{ggd3} becomes
\begin{equation}\label{ggd4}
	\tilde{r}_{g}
	=\tilde{r}_{\gamma}^{~}
	+\frac{1+z_{A}^{~}}{1+(1+\Delta z)z_{A}^{~}}\frac{W_{5}}{H_{B}^{~}\sqrt{\Omega_{\Lambda}^{~}}} {~},
\end{equation}
where
\begin{eqnarray}
	W_{5}^{~}&\!\equiv\!&
	\Big(1+\frac{z_{A}^{~}\Delta z}{1+z^{~}_{A}}\Big)	\,_{2}^{~}F_{1}^{~}\!\Big[\frac{1}{3},\frac{1}{2};\frac{4}{3};-\frac{\Omega_{m}^{~}}{\Omega_{\Lambda}^{~}}\Big(1+\frac{z_{A}^{~}\Delta z}{1+z^{~}_{A}}\Big)^{3}_{~}\Big]\nonumber\\	&\!~\!&-~_{2}^{~}F_{1}^{~}\!\Big(\frac{1}{3},\frac{1}{2};\frac{4}{3};-\frac{\Omega_{m}^{~}}{\Omega_{\Lambda}^{~}}\Big).
\end{eqnarray}
\begin{figure}[!htb]
\center{
\subfigure[]{\includegraphics[width=4cm]{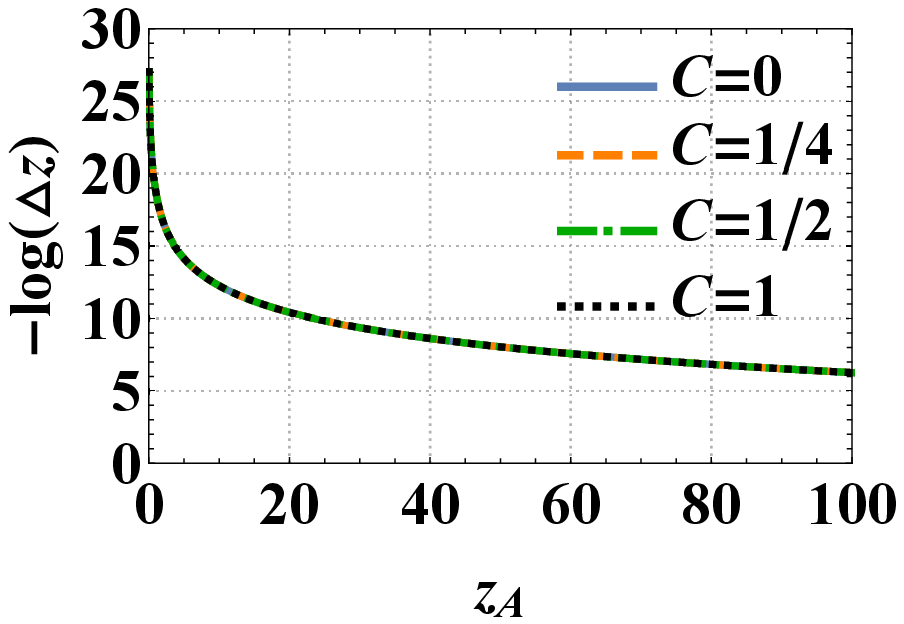}\label{constraint21}}
\subfigure[]{\includegraphics[width=4.1cm]{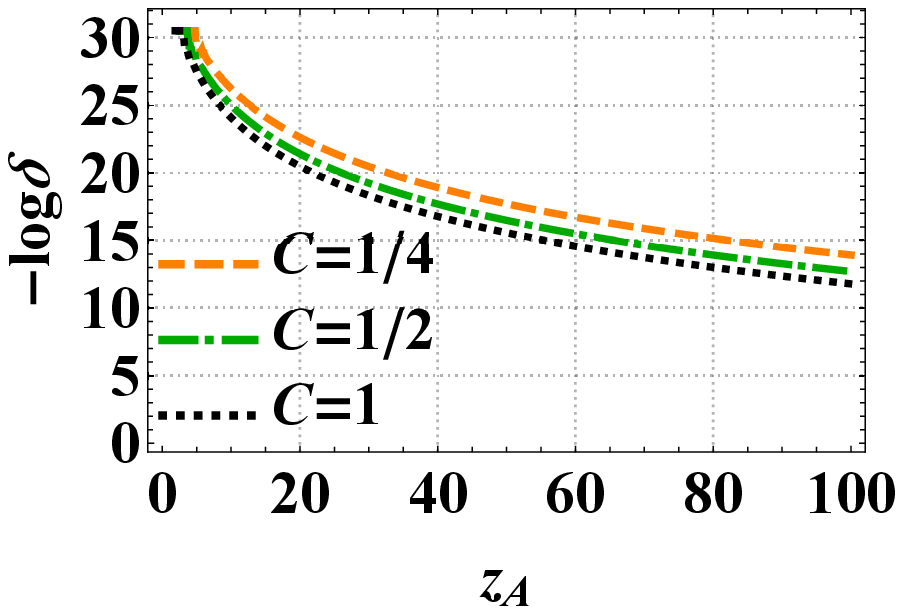}\label{constraint22}}
}
\caption{The deviation between the source redshifts given by the EMW observation ($z'_{A}$) and GW observation ($z_{A}^{~}$). (a) The relative deviation $\Delta z$ with respect to the source redshift $z_{A}^{~}$. The parameter $C$ is set to $C=0$ (blue solid curve), $C=1/4$ (orange dashed curve), $C=1/2$ (green dashed and dotted curve), and $C=1$ (dark dotted curve). (b) The parameter $\delta$ with respect to the source redshift $z_{A}^{~}$. The parameter $C$ is set to $C=1/4$ (orange dashed curve), $C=1/2$ (green dashed and dotted curve), and $C=1$ (black dotted curve).}
\label{constraint2}
\end{figure}

In Fig.~\ref{constraint21}, we show the evolution of the relative deviation $\Delta z$ with respective to the source redshift $z_{A}^{~}$. The source redshift we consider ranges from $z_{A}^{~}=0.01$ to $z_{A}^{~}=100$. It can be seen that the relative deviation $\Delta z$ is extremely small when $z_{A}^{~}$ is closed to the lower boundary, which is consistent with the result (i.e.,  the time delay is extremely short for a low-redshift source) in the joint observations of GW170817/GRB 170817A. The relative deviation $\Delta z$ increases with the source redshift $z_{A}^{~}$, which is also consistent with our previous conclusion that the time delay between the detections of the GW and EMW signals increases with the source redshift. Moreover, it is found that when the source redshift approaches $z_{A}^{~}\sim23$, the difference between $z'_{A}$ and $z_{A}^{~}$ is of the order $z'_{A}-z_{A}^{~}\sim0.001$. When the source redshift reaches $z_{A}^{~}\sim44$, the difference becomes significant ($z'_{A}-z_{A}^{~}\sim0.01$). Therefore, for the future joint observations, if the $\text{AdS}_{6}^{~}$ radius is about $\ell^{2}_{~}\approx3.84\,\text{Mpc}^{2}_{~}$, the EM counterpart of a high-redshift GW event can not reach the observer within a reasonable observation time. If we expect that in the future observations, both a GW signal with a high-redshift source and its EM counterparts could be detected within the observation time, the $\text{AdS}_{6}^{~}$ radius must be limited to a smaller range. It will be discussed in the next section.

Note that, in Fig.~\ref{constraint21}, the four curves are overlapped with each other. This is because the contribution of the brane's motion $\Psi(t)$ to the relative deviation $\Delta z$ is tiny under the low-energy approximation. To show this slight difference for different $C$, we introduce the following parameter:
\begin{equation}
	\delta\equiv\Delta z-\Delta z_{0}^{~},
\end{equation}
where $\Delta z_{0}^{~}$ is the value of the relative deviation $\Delta z$ at $C=0$. The behavior of $\delta$ with respect to the source redshift $z_{A}^{~}$ is plotted in Fig.~\ref{constraint22}. Therein, we use the black dotted curve with $C=1$ to show the upper boundary of $\delta$. One can find that the parameter $\delta$ increases with the source redshift $z_{A}^{~}$ but the increasing rate is very slow. { So according to our calculations and analysis, it is not expected to detect the new physics from the model we study in the next generation of GW observations ($z_{A}^{~}\lesssim100$).

\begin{figure}[!htb]
\center{
\includegraphics[width=4cm]{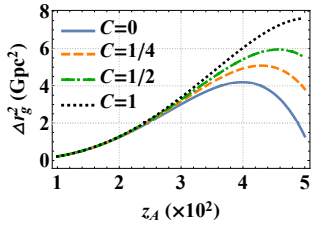}
}
\caption{{ High-order correction to the gravitational horizon radius. The parameter $C$ is set to $C=0$ (blue solid curve), $C=1/4$ (orange dashed curve), $C=1/2$ (green dashed and dotted curve), and $C=1$ (dark dotted curve), respectively.}}
\label{constraint4}
\end{figure}
Comparing the last two terms in the gravitational horizon radius~\eqref{ghr6}, we find the contribution from the brane's motion $\Psi(t)$ is significant when $\ell^{2}_{~}H_{B}^{2}z_{A}^{2}\sim1$. In Fig.~\ref{constraint4}, we plot the high-order correction to the gravitational horizon radius under the low-energy approximation, where we define
\begin{eqnarray}\label{hocorrection1}
	\Delta r_{g}^{2}\equiv r_{g}^{2}-r_{\gamma}^{2}
	\!&\!\approx\!&\! \frac{1}{H_{B}^{2}}\bigg[
	\ell^{2}_{~}H_{B}^{2}(W_{1}^{~}W_{2}^{~}-z_{A}^{2})\nonumber\\
	\!&\!~\!&\!
	+\frac{\ell^{4}_{~}H_{B}^{4}\Omega_{\Lambda}^{~}}{4(1+C^{2}_{~})}(W^{2}_{2}-W^{~}_{1}W^{~}_{3})\bigg].
\end{eqnarray}
It is shown that the correction will decrease finally in all cases, because the last term in~\eqref{hocorrection1} is always negative for low redshift and becomes significant when the source redshift approaches to $z_{A}^{~}\sim \ell^{-2}_{~}H_{B}^{-2}$. Besides, we find a nonvanishing $C$ can suppress the contribution from the last term. Consequently, the brane's motion $\Psi(t)$ allows a larger gravitational horizon radius. However, we should note that the high-order terms beyond $\mathcal{O}(\ell^{4}_{~}H_{B}^{4})$ in the gravitational horizon radius~\eqref{ghr6} and therefore in~\eqref{hocorrection1}, might be significant as well when $z_{A}^{~}\sim \ell^{-2}_{~}H_{B}^{-2}$. So it requires a deeper analysis when $\ell^{2}_{~}H_{B}^{2}z_{A}^{2}\sim1$, which is not referred in this paper.
}
%Therefore, the change in $\delta$ with the source redshift is negligible until the source redshift is extremely large. Moreover, since we have concluded that the contribution of the brane's motion $\Psi(t)$ to the relative deviation $\Delta z$ is tiny, it is an expected result that the distinction between the three curves is not obvious, which is also consistent with the results in Fig.~\ref{constraint21}.

\subsection{Constraint for future observations}

The detections of GWs originated from the coalescence of compact binaries by LIGO and Virgo detectors help us to advance the understanding of astrophysics~\cite{Abbott5}, fundamental physics~\cite{Abbott6} and cosmology~\cite{Abbott7}. For the Advanced LIGO and Advanced Virgo, they are designed to observe the GWs with the frequency ranging from $0.1\,\text{kHz}$ to $1\,\text{kHz}$~\cite{Harry1,Aasi1,Acernese1}. The Einstein Telescope (ET) is expected to have a wider sensitivity band and a smaller strain noise spectrum than the second generation~\cite{Broeck1,Punturo1}. As a ground-based detector, the ET is still not sensitive to the GWs below $1\,\text{Hz}$, and the signals observed by it can not last more than 9 days. Unlike the ground-based detectors, space-based detectors are able to detect low-frequency GWs and have a longer observation time. For the Laser Interferometer Space Antenna (LISA), Taiji, and TianQin, their sensitivity bands could cover the frequency from $0.1\,\text{mHz}$ to $0.1\,\text{Hz}$~\cite{Danzmann1,Seoane1,Luo1,Hu1}. The Deci-hertz Interferometer Gravitational Wave Observatory (DECIGO) has an optimal band from $0.1\,\text{Hz}$ to $10\,\text{Hz}$, which builds a bridge between the space-based detectors and ground-based detectors~\cite{Seto1,Kawamura1,Kawamura2}. As for the nanohertz GWs, they are expected to be observed by pulsar timing arrays (PTAs)~\cite{Kramer1,Jenet1,Hobbs1,Hobbs2,Hobbs3}. For other proposed subjects, one can refer to Ref.~\cite{Gong1} and references therein.

\begin{figure*}[!htb]
\center{
\subfigure[]{\includegraphics[width=4cm]{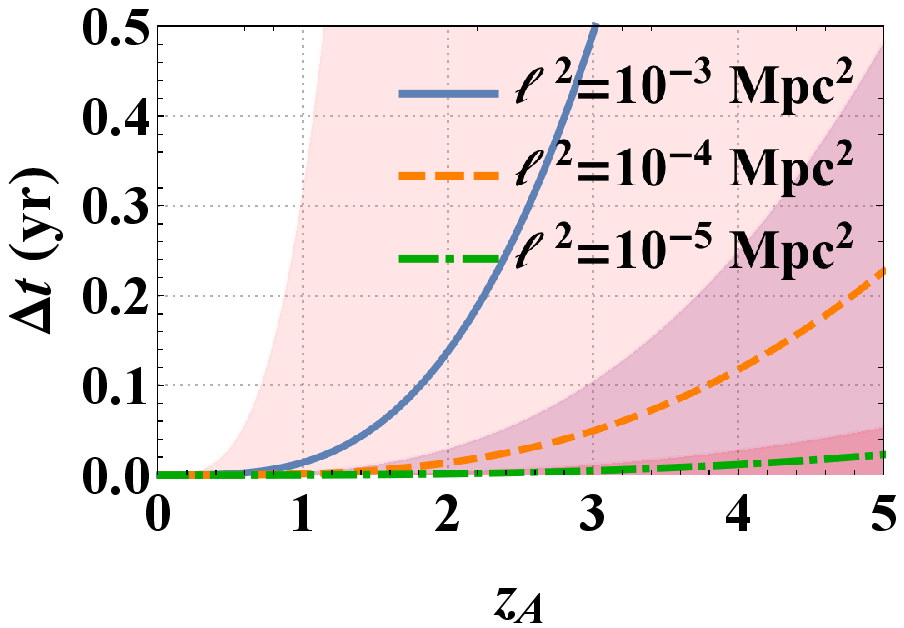}\label{constraint31}}
\quad
\subfigure[]{\includegraphics[width=4.1cm]{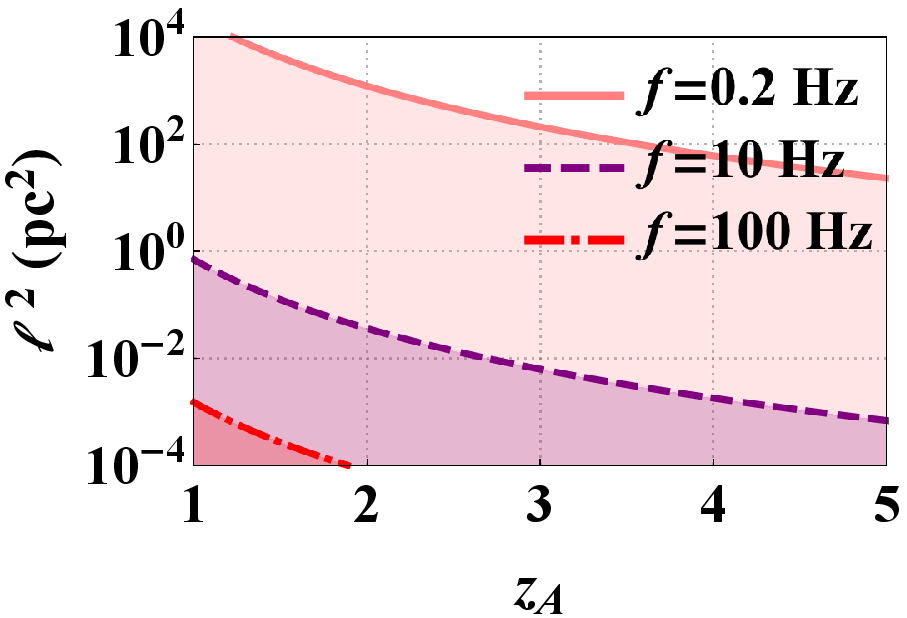}\label{constraint32}}
\quad
\subfigure[]{\includegraphics[width=4.7cm]{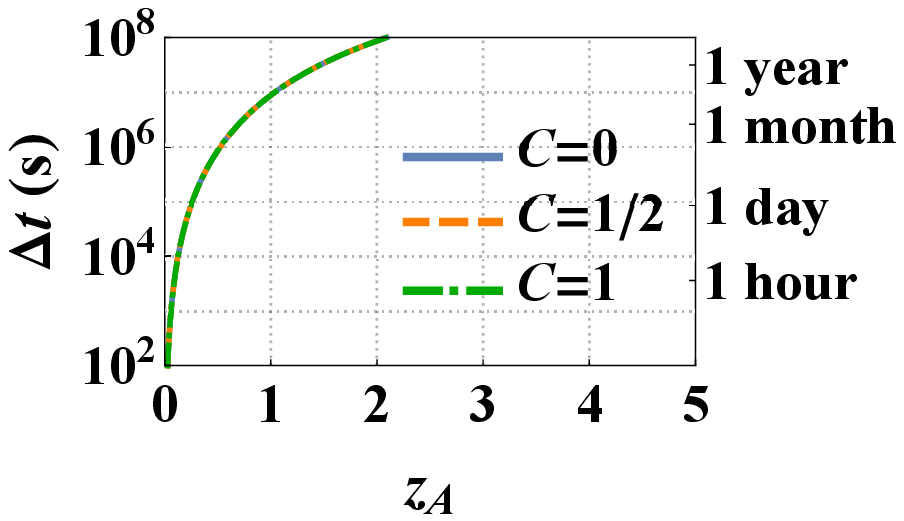}\label{constraint33}}
}
\caption{The time delay and allowed $\text{AdS}_{6}^{~}$ radius based on the DECIGO/BBO. (a) The predicted time delay with respect to the source redshift. The $\text{AdS}_{6}^{~}$ radius is set as $\ell^{2}_{~}=10^{-3}_{~}\,\text{Mpc}^{2}_{~}$ (blue solid curve), $\ell^{2}_{~}=10^{-4}_{~}\,\text{Mpc}^{2}_{~}$ (orange dashed curve), and $\ell^{2}_{~}=10^{-5}_{~}\,\text{Mpc}^{2}_{~}$ (green dashed and dotted curve). (b) The allowed $\text{AdS}_{6}^{~}$ radius with respect to the source redshift. The observed GW frequency is set to $f=0.2\,\text{Hz}$ (pink solid curve), $f=10\,\text{Hz}$ (purple dashed curve), and $f=100\,\text{Hz}$ (red dashed and dotted curve). In (a) and (b), we set $C=0$. (c) The contribution of the brane's motion $\Psi(t)$ to the time delay. The parameter $C$ is set as $C=0$ (blue solid curve), $C=1/2$ (orange dashed curve), and $C=1$ (green dashed and dotted curve). The $\text{AdS}_{6}^{~}$ radius is set as $\ell^{2}_{~}=0.02\,\text{Mpc}^{2}_{~}$. In (a) and (c), we set $f=0.2\,\text{Hz}$.}
\label{constraint3}
\end{figure*}

In this section, we focus on GWs and their EM counterparts originated from BNS's with a high-redshift. We set $m=1.4\,M_{\odot}^{~}$ for each of the BNS components. For the DECIGO and Big Bang Observer (BBO), the GW from this BNS could enter their bands~\cite{Yagi1}. Taking into account the upper frequency cutoff of the binary white dwarf, we can set a lower boundary on the GW frequency as $f=0.2\,\text{Hz}$. The upper boundary is optimistically chosen as $f=100\,\text{Hz}$. For the DECIGO/BBO, to detect the GW from a $(1.4+1.4)\,M_{\odot}^{~}$ BNS, the corresponding source redshift can not exceed $z=5$. The effective observation time of the GW also depends on the source redshift. When the source redshift approaches to $z\sim0.02$, the GW can stay on the detectors for almost $1\,\text{yr}$, whereas the observation time is reduced to nearly one month when $z\sim4$. If the time delay between the GW and its EM counterpart could be found on the DECIGO/BBO, the EM counterpart has to reach the detectors during the observation time. Based on the expectation, one could calculate the upper limits to the time delay for different source redshifts. As a result, those upper limits would finally lead to a stronger constraint on the $\text{AdS}_{6}^{~}$ radius. The finial results are shown in Fig.~\ref{constraint3}.

In Fig.~\ref{constraint31}, the pink, purple, and red regions correspond to $\ell^{2}_{~}\leq 0.02\,\text{Mpc}^{2}_{~}$, $\ell^{2}_{~}\leq 211.87\,\text{pc}^{2}_{~}$, and $\ell^{2}_{~}\leq 23.30\,\text{pc}^{2}_{~}$, respectively. These values of the $\text{AdS}_{6}^{~}$ radius are chosen on account of the assumption that the EM counterpart could be observed during the DECIGO/BBO observation time for $z_{A}^{~}=1$, $z_{A}^{~}=3$, and $z_{A}^{~}=5$, respectively. Note that the observation time is also affected by the GW frequency. For a BNS with the redshift $z=1$, the GW can stay on the DECIGO/BBO for at most four months. If the minimum GW frequency is around $18\,\text{Hz}$, the observation time will be reduced to only one minute. Therefore, once the EM counterpart of such a GW signal is expected to be found by the follow-up observations, the $\text{AdS}_{6}^{~}$ radius has to satisfy $\ell^{2}_{~}\leq0.14\,\text{pc}^{2}_{~}$. In Fig.~\ref{constraint32}, we use the colored regions to denote the allowed $\text{AdS}_{6}^{~}$ radius for the DECIGO/BBO. The pink, purple, and red regions represent the allowed $\text{AdS}_{6}^{~}$ radius calculated by the EM counterpart that comes from a GW signal with the minimum frequency $0.2\,\text{Hz}$, $10\,\text{Hz}$, and $100\,\text{Hz}$, respectively. The red region gives very strong constraints on the $\text{AdS}_{6}^{~}$ radius. Eventually, we obtain the strongest constraint on the $\text{AdS}_{6}^{~}$ radius as $\ell^{2}_{~}\lesssim0.02\,\text{Mpc}^{2}_{~}$ with $z_{A}^{~}=1$ and $f=0.2\,\text{Hz}$ on the DECIGO/BBO. Moreover, our result shows that the contribution of brane's motion $\Psi(t)$ on the time delay is still not significant (see Fig.~\ref{constraint33}).

\section{Conclusion}\label{sec6}

The braneworld theory allows the higher-dimensional null geodesic to deviate from the trajectory of a light confined on the brane. If such deviation exists, the trajectory of a GW signal that causally connects the source and the observer could be a shorter path than the path of a light. It provides an opportunity to find the clues of extra dimensions through the joint GW and EMW observations on the brane. Assuming that there is a source emitting a GW signal and an EMW signal simultaneously, an observer located on the brane will detect these two signals successively. And the time delay between the arrivals of the signals can be measured by the joint GW and EMW observations. In the paper, we used such a property of GWs in the braneworld theory to investigate the structure of extra dimensions.

We considered a six-dimensional static spacetime with a bulk cosmological constant. %Solving the field equations, the metric~\eqref{6metric2} describing a six-dimensional AdS spacetime was found. We then embedded a 4-brane in this background spacetime, with the induced metric being the FLRW metric~\eqref{4metric2}.
The universe is regarded as a 4-brane embedded the background spacetime. The brane's back-reaction to the background spacetime is ignored in the paper, and the brane's motion in the bulk is described by $\mathcal{R}(t)$ and $\Psi(t)$. %In addition, by comparing the bulk metric and the induced metric, a relation~\eqref{tT1} between the bulk time and cosmic time was found. Focusing on a six-dimensional null geodesic with $d\theta=0$ and $d\phi=0$, we found three Killing vectors defined on it. They are relative to three constants of motion, giving the corresponding equations of motion~\eqref{eom1} of the geodesic.
We derived the gravitational horizon radius~\eqref{ghr42} under the low-energy approximation $\ell H\ll1$. It is the projection of the trajectory of a six-dimensional null geodesic on the 4-brane for a given time interval. %It can be different from the photon horizon radius because of the shortcut. And the discrepancy between these two radii is related to the time delay in the joint observations.
%So we converted the expressions of them in terms of the observables, see Eqs.~\eqref{ghr5} and~\eqref{phr1}, under the assumption that the $\Lambda$CDM model can well describe the late-time universe. Remarkably,
It was found that the gravitational horizon radius recovers the photon horizon radius at the leading order. The contribution of extra dimensions is manifest in higher-order terms. And the gravitational horizon radius has the same form as the one derived in the five-dimensional model~\cite{Visinelli1}, when the brane's motion $\Psi(t)$ vanishes.

In 2017, the LIGO/Virgo detectors reported a GW event (GW170817) originated from a BNS system $40^{+8}_{-14}\,\text{Mpc}$ from the earth~\cite{Abbott2}. Subsequently, the EMW observation found a sGRB signal (GRB 170817) which was $1.74^{+0.05}_{-0.05}\,\text{s}$ later than the detection of GW170817~\cite{Coulter1,Goldstein1,Savchenko1,Pan1}. These two signals were soon proved to be emitted by the same source located in NGC 4993~\cite{Coulter1,Pan1}. Since the source redshift $z_{A}^{~}=0.008^{+0.002}_{-0.003}$ of GW170817/GRB 170817A is low enough, we can directly use the time delay to constrain the scale of the $\text{AdS}_{6}^{~}$ radius through the relation~\eqref{ggd2}. Note that this relation was deduced under the assumption that the two signals were emitted simultaneously, while astrophysics models allow a time lag $(-100\,\text{s},1000\,\text{s})$ between the emissions of GW170817/GRB 170817A~\cite{Shibata1,Rezzolla1,Paschalidis1,Rezzolla2,Ciolfi1,Tsang1}. Therefore, we revised the time delay between the detections of the two signals to $(0\,\text{s},101.74\,\text{s})$. Our result shows that the upper limit to the $\text{AdS}_{6}^{~}$ radius is about $\ell^{2}_{~}\approx 3.84\,\text{Mpc}^{2}_{~}$, where the parameter $C$ is set to zero.
%As was discussed before (see also Refs.~\cite{Abdalla1,Abdalla2,Abdalla3,Cuadros-Melgar1}), if we do not concern the brane's motion $\Psi(t)$, the gravitational horizon radius recovers the one given in Ref.~\cite{Visinelli1}.
%Thus this upper limit also constrains the $\text{AdS}_{5}^{~}$ radius. Note that our constraint is looser than the one, $\ell\lesssim0.535\,\text{Mpc}$, given in Ref.~\cite{Visinelli1}. This is because we included contributions from astrophysics models.
We pointed out that the brane's motion $\Psi(t)$ contributes little to the shortcut of the GW for GW170817/GRB 170817A, since the source redshift is very low.

For the future joint multi-messenger observations, we expect that most of the target compact binary coalescences have high redshifts. In this case, the relation~\eqref{ggd2} under the low-redshift approximation may be not valid for these events. Therefore, we derived an extension~\eqref{ggd3} of~\eqref{ggd2} and converted the time delay into a discrepancy ($\Delta z$) between the redshifts given by the GW observation ($z_{A}^{~}$) and the EMW observation ($z'_{A}$), respectively. For a given $\text{AdS}_{6}^{~}$ radius ($\ell^{2}_{~}\approx3.84\,\text{Mpc}^{2}_{~}$), we found that the redshift discrepancy $\Delta z=z'_{A}-z_{A}^{~}$ increases with the source redshift $z_{A}^{~}$ given by the GW observation. Our result shows that the redshift discrepancy is of the order $\Delta z\sim0.001$ when $z_{A}^{~}$ is about $z_{A}^{~}\sim20$, and it becomes significant ($\Delta z\sim0.01$) when the source redshift approaches $z_{A}^{~}\sim40$. Such high redshift differences indicate that if the $\text{AdS}_{6}^{~}$ radius is $\ell^{2}_{~}\approx3.84\,\text{Mpc}^{2}_{~}$ in the model, the EM counterparts will never be found within a reasonable observation time. If we expect to detect the EM counterpart of a high-redshift GW event within a reasonable observation time, a stronger constraint on the $\text{AdS}_{6}^{~}$ radius is required.

At the end of this paper, we considered a GW signal from a $(1.4+1.4)\,M_{\odot}^{~}$ BNS, which could enter the sensitivity bands of DECIGO and BBO. We  also assumed that the frequency of the GW is $0.2\,\text{Hz}$, and that the GW is triggered by the BNS at $z_{A}^{~}=1$. For the sake of simplicity, we focused on the simultaneously triggered GW and EM counterpart. And the astrophysics influence to their emissions was ignored. We then found that, to detect the EM counterpart within the observation time of DECIGO/BBO, the $\text{AdS}_{6}^{~}$ radius must be limited to $\ell^{2}_{~}\lesssim0.02\,\text{Mpc}^{2}_{~}$. It is a stronger constraint than the previous one obtained by GW170817/GRB 170817A. Moreover, our result also shows that the brane's motion $\Psi(t)$ does not give a significant contribution to the time delay between the detections of GW signal and its EM counterpart for the upper limit of the redshift that the DECIGO/BBO can detect.

\section*{Acknowledgements}

%We thank Si-Jiang Yang, Hai Yu, and Xiang-Ru Li for useful discussions. We also thank the referee for his/her helpful suggestions.
This work was supported by the National Key Research and Development Program of China (Grant No.~2020YFC2201503), the National Natural Science Foundation of China (Grants No.~11875151 and No.~12047501), the 111 Project (Grant No.~B20063), the Fundamental Research Funds for the Central Universities (Grants No.~lzujbky-2020-it04), and ``Lanzhou City's scientific research funding subsidy to Lanzhou University'' (Grant No.~2021CXZX-012).

\end{document}